\documentclass[amsmath,prx,twocolumn,superscriptaddress]{revtex4}
\usepackage{color}
\usepackage{amsfonts}
\usepackage{amssymb}
\usepackage{graphicx}
\usepackage{pstricks}

\begin{document}

\title{Entropy and information flow in quantum systems strongly coupled to baths}
\author{Nikhil Seshadri}
\affiliation{University City High School,  San Diego, CA 92122, USA}
\author{Michael Galperin}
\email{migalperin@ucsd.edu}
\affiliation{Department of Chemistry \& Biochemistry, University of California San Diego, La Jolla, CA 92093, USA}

\begin{abstract}
Considering von Neumann expression for reduced density matrix as thermodynamic entropy of 
a system strongly coupled to baths,
we use nonequilibrium Green's function (NEGF) techniques to derive bath and energy resolved expressions for
entropy, entropy production, and information flows. The consideration is consistent with dynamic
(quantum transport) description and expressions reduce to expected forms in limiting cases of
weak coupling or steady-state. Formulation of the flows in terms of only system degrees freedom 
is convenient for simulation of thermodynamic characteristics of open nonequilibrium quantum systems.
We utilize standard NEGF for derivations in noninteracting systems, Hubbard NEGF is used for
interacting systems. Theoretical derivations are illustrated with numerical simulations
within generic junction models. 
\end{abstract}
\maketitle
\section{Introduction}\label{intro}
Rapid development of experimental techniques at the nanoscale in the last decade
has led to miniaturization of devices for energy storage and conversion to sizes where quantum mechanics
becomes relevant.
For example, thermoelectric single atom and single molecule junctions are expected to operate more effectively 
compared to their bulk analogs due to possible utilization of quantum 
effects~\cite{reddy_thermoelectricity_2007,lee_heat_2013,kim_electrostatic_2014,zotti_heat_2014, cui_perspective:_2017,cui_quantized_2017,cui_peltier_2018,cui_thermal_2019}.
Proper description of performance of nanoscale devices for energy conversion requires development
of corresponding nonequilibrium quantum thermodynamics. Moreover, with molecules
chemisorbed on at least one of the contacts, thermodynamic theory should account for strong
system-bath couplings.

In recent years there is a surge of research in this field both experimentally~\cite{jezouin_quantum_2013,pekola_towards_2015,hartman_direct_2018,klatzow_experimental_2019} 
and theoretically~\cite{esposito_quantum_2015,esposito_nature_2015,uzdin_quantum_2016,katz_quantum_2016,carrega_energy_2016,strasberg_nonequilibrium_2016,kato_quantum_2016,hsiang_quantum_2018,perarnau-llobet_strong_2018,funo_path_2018,strasberg_repeated_2019}.
One of the guiding principles for theoretical research is consistency of the thermodynamic description
with underlying system dynamics~\cite{kosloff_quantum_2013,alicki_introduction_2018,kosloff_quantum_2019}.
Thermodynamic formulations for systems strongly coupled to their baths can be roughly divided into
two groups: 1.~supersystem-superbath and 2.~system-bath approaches.

The first group complements the physical supersystem (system strongly coupled to its baths) with a set of superbaths 
(additional baths) weakly coupled to the supersystem. Choosing thermodynamic border at the weak link 
(i.e. between the supersystem and superbath) and utilizing tools of standard thermodynamics, 
thermodynamics of the system strongly coupled to its baths is introduced as the difference between 
thermodynamic formulations for the supersystem and free baths (both weakly coupled to the superbaths).
This approach was pioneered in Refs.~\onlinecite{hanggi_finite_2008, ingold_specific_2009}
and extended to nonequilibrium using scattering theory~\cite{ludovico_dynamical_2014,ludovico_periodic_2016,ludovico_dynamics_2016,bruch_landauer-buttiker_2018,semenov_transport_2020},
density matrix~\cite{dou_universal_2018,oz_evaluation_2019,dou_universal_2020,oz_numerical_2020,rivas_strong_2020,brenes_tensor-network_2020}, and 
nonequilibrium Green's function~\cite{bruch_quantum_2016,ochoa_quantum_2018,haughian_quantum_2018} formulations.
However, this way of building the thermodynamic description is inconsistent with the microscopic dynamical description of 
the physical system~\cite{bergmann_greens_2020}. Thus, it is natural that the approach has difficulties in describing
energy fluctuations~\cite{ochoa_energy_2016}.

The second group builds the thermodynamic description using physical supersystem only (system strongly
coupled to its baths, no additional baths are assumed). This approach postulates 
von Neumann entropy expression for reduced density matrix of the system as 
thermodynamic entropy. Then, the second law is formulated in terms of system and baths characteristics.
The integral expression for entropy production defined as relative entropy is guaranteed to be positive,
although it does not increase monotonically. The approach was originally proposed in 
Refs.~\onlinecite{lindblad_non-equilibrium_1983,peres_quantum_2002,esposito_entropy_2010} and later used in a number of 
studies~\cite{sagawa_second_2012,kato_quantum_2016,strasberg_quantum_2017}.
An attractive feature of the formulation is  its consistency with the dynamical (quantum transport) 
description~\cite{bergmann_greens_2020}.
Although this approach was recently criticized as being formulated using external to the system (i.e. baths)
variables and as being not consistent with expected behavior at thermal equilibrium~\cite{rivas_strong_2020},
it seems, that Green's function based formulation presented below is capable of satisfying the
requirements: below we express all thermodynamic characteristics of the system in terms of system variables only, 
and reaching thermal equilibrium was shown in Ref.~\onlinecite{sieberer_thermodynamic_2015}
to be consequence of a symmetry of the Schwinger-Keldysh action.  

Another active area of research establishes a connection between thermodynamic and information theory.
A number of experimental~\cite{berut_experimental_2012,cottet_observing_2017,naghiloo_information_2018,masuyama_information--work_2018}
and theoretical~\cite{bhattacharya_thermodynamical_2013,horodecki_fundamental_2013,brandao_resource_2013,parrondo_thermodynamics_2015,brandao_second_2015,gour_resource_2015,sharma_landauer_2015,goold_role_2016,vinjanampathy_quantum_2016,gagatsos_entropy_2016,strasberg_quantum_2017,sanchez_autonomous_2019}
studies establish foundations for thermodynamics of quantum information.
Recently, discussion of quantum information flows in Markovian open quantum systems was presented
in Refs.~\onlinecite{horowitz_thermodynamics_2014,ptaszynski_thermodynamics_2019}.
In particular, the local Clausius inequality relating entropy balance of a subsystem to the information
flow was established for system weakly coupled to its baths (Markov dynamics with second order
in system-bath couplings).

Here, we utilize the system-bath approach to extend description of quantum information flow to 
non-Markov dynamics of systems strongly coupled to their baths. We reformulate the
system-bath approach in terms of Green's functions. This leads to local (system variables only)
formulation and allows to introduce separability of baths contributions beyond weak (second order) coupling.
Also, strong coupling results in energy-resolved expressions for entropy and information fluxes.
The structure of the paper is as follows: in Section~\ref{theory} we introduce Green's function based formulation
first for non-interacting and then for interacting systems. 
The former uses the standard nonequilibrium Green's function (NEGF) technique~\cite{haug_quantum_2008,balzer_nonequilibrium_2013,stefanucci_nonequilibrium_2013}, 
while the latter employs its many-body flavor - the Hubbard NEGF~\cite{chen_nonequilibrium_2017,miwa_towards_2017,bergmann_electron_2019}.
Numerical simulations within generic junction models for thermodynamic and information 
properties of open quantum systems are presented in Section~\ref{numres}.
Section~\ref{conclude} summarizes our findings and indicates directions for future research.


\begin{figure}[b]
\centering\includegraphics[width=0.8\linewidth]{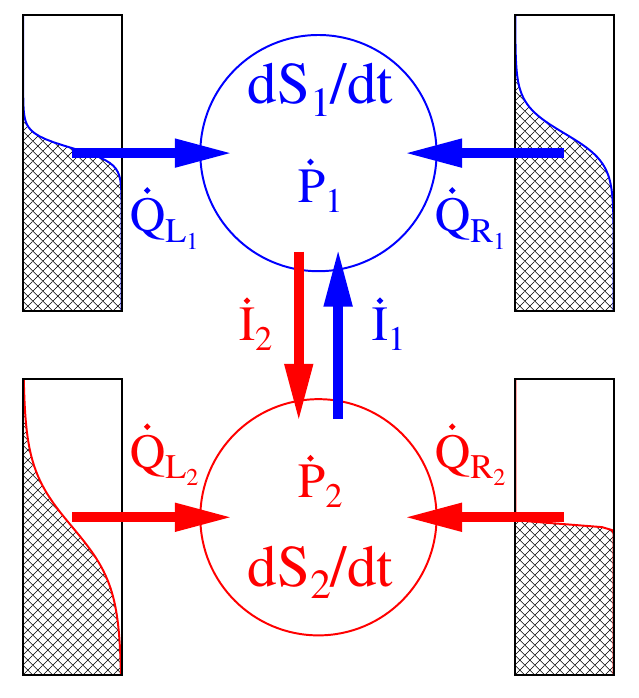}
\caption{\label{fig1}
Sketch of a junction which consists of two sub-systems, each coupled to its own baths.
Entropy change $dS_i/dt$ in subsystem $i$ is caused by heat fluxes $\dot{Q}_{L_i}$ and $\dot{Q}_{R_i}$,
entropy production rate $\dot{P}_i$, and information flow $\dot{I}_i$.
}
\end{figure}


\section{\label{theory}Thermodynamics: Green's functions formulation}
We consider a junction which consists of system $S$ strongly coupled to set of baths $\{B\}$. The system and 
the couplings are subjected to time-dependent driving. 
Hamiltonian of the model is (here and below $e=k_B=\hbar=1$)
\begin{equation}
\label{Hdef}
\hat H(t) = \hat H_S(t) + \sum_B \bigg(\hat H_B + \hat V_{SB}(t)\bigg)
\end{equation} 
Before system-bath coupling is established at time $t_0$, baths are assumed to be in thermal equilibrium
characterized by temperature $\beta_B=1/T_B$ and chemical potential $\mu_B$.

System-bath approach to thermodynamics introduced in Ref.~\onlinecite{esposito_entropy_2010} defines
von Neumann expression as the system entropy
\begin{equation}
\label{Sdef}
 S(t) \equiv - \mbox{Tr}_S \big[\hat\rho_{S}(t)\,\log\hat\rho_{S}(t)\big]
 \end{equation}
Here, $\hat\rho(t)$ is the density operator of the universe (system plus baths)
and $\hat\rho_S(t)\equiv \mbox{Tr}_{\{B\}}\big[\hat\rho(t)\big]$.
Considering the system and baths initially in the product state 
$\hat\rho(t_0)=\hat\rho_S(t_0)\otimes_B \hat\rho_B^{eq}$,
this leads to integral version of the second law in the form~\cite{esposito_entropy_2010}
\begin{equation}
\label{law2}
 \Delta S(t) = \sum_B \beta_B\, Q_B(t) + \Delta P(t) 
\end{equation}
where
\begin{equation}
\label{QBSi}
\begin{split}
Q_B(t) &\equiv -\mbox{Tr}\big[\hat H_B\,\hat\rho(t)\big] + \mbox{Tr}\big[\hat H_B\,\hat\rho(t_0)\big] 
\\
 \Delta P(t) &\equiv D\big[\hat\rho(t)\vert\vert \hat\rho_S(t)\otimes_B\hat\rho_B^{eq}\big]
 \end{split} 
\end{equation}
are the heat transferred from bath B into the system and entropy production during time $t-t_0$.
Here, $D[\hat\rho_1\vert\vert\hat\rho_2] = \mbox{Tr}[\hat\rho_1\log\hat\rho_1] - \mbox{Tr}[\hat\rho_1\log\hat\rho_2]$
is the quantum relative entropy.

Below, starting from (\ref{Sdef}) and using NEGF techniques we introduce expressions
for thermodynamic characteristics (entropy, entropy production, heat and information fluxes)
in terms of single-particle Green's functions. These expressions are defined in the system subspace only
and are suitable for actual calculations.
Note that expression for heat in (\ref{QBSi}) is consistent with definitions of particle and energy fluxes accepted in 
quantum transport considerations. Also, the system-bath approach (\ref{Sdef})-(\ref{QBSi}) does not introduce artificial
additions to the physical system. Thus, thermodynamic formulation is consistent with dynamic
(quantum transport) description. 
Note also that while thermodynamic formulation of Ref.~\onlinecite{esposito_entropy_2010}
strict in assumption of initial product state, Green's function based analysis in principle allows to
relax this restriction by shifting consideration from the Keldysh to Konstantinov-Perel contour~\cite{konstantinov_diagram_1961,wagner_expansions_1991}.
First, we utilize standard NEGF~\cite{haug_quantum_2008,balzer_nonequilibrium_2013,stefanucci_nonequilibrium_2013} 
and consider non-interacting system.
After that, we use the Hubbard NEGF~\cite{chen_nonequilibrium_2017,miwa_towards_2017,bergmann_electron_2019} 
to generalize the consideration to interacting systems.

Exact formulations of the second law of thermodynamics in the form of energy resolved partial Clausius, Eq.(\ref{dlaw2}),
and local Clausius, Eq.~(\ref{Clausius_loc}), expressions together with equations for the 
energy resolved entropy, heat, entropy production and information fluxes,
Eqs.~(\ref{SQPE_nonint}), (\ref{dSpE_nonint}), (\ref{dIpE_nonint}), (\ref{SQPE}), (\ref{dSpE}), and (\ref{dIpE}),
 are the main results of our consideration. 

\subsection{Non-interacting system}
First, we consider an open non-interacting Fermi system described by the Hamiltonian (\ref{Hdef}) with
\begin{equation}
\label{Hnonint}
\begin{split}
 \hat H_S(t) &= \sum_{i,j\in S} H^{S}_{ij}(t)\,\hat d_i^\dagger\hat d_j,\qquad 
 \hat H_B = \sum_{k\in B} \varepsilon_k\hat c_k^\dagger\hat c_k
 \\
 \hat V_{SB}(t) &= \sum_{i\in S}\sum_{k\in B}
 \bigg( V_{ik}(t)\,\hat d_i^\dagger\hat c_k + V_{ki}(t)\,\hat c_k^\dagger\hat d_i\bigg)
\end{split}
\end{equation}
Here, $\hat d_i^\dagger$ ($\hat d_i$) and $\hat c_k^\dagger$ ($\hat c_k$) creates (annihilates)
electron in orbital $i$ of the system and state $k$ of the bath $B$, respectively.

For the non-interacting system  (\ref{Hnonint}) entropy (\ref{Sdef}) is~\cite{sharma_landauer_2015,bergmann_greens_2020}
\begin{equation}
\label{Snonint}
\begin{split}
 S(t) =& -\mbox{Tr}_S\bigg[-i\mathbf{G}^{<}(t,t)\,\ln\big(-i\mathbf{G}^{<}(t,t)\big)\bigg]
 \\ &
 -\mbox{Tr}_S\bigg[i\mathbf{G}^{>}(t,t)\,\ln\big(i\mathbf{G}^{>}(t,t)\big)\bigg]
 \end{split}
\end{equation}
where $\mathbf{G}^{\gtrless}(t,t)$ are matrices in the system subspace representing 
greater/lesser projections of the single-particle Green's function defined on the Keldysh contour 
as~\cite{haug_quantum_2008,balzer_nonequilibrium_2013,stefanucci_nonequilibrium_2013}
\begin{equation}
\label{Gdef}
G_{ij}(\tau,\tau') \equiv -i\langle T_c\,\hat d_i(\tau)\,\hat d_j^\dagger(\tau')\rangle
\end{equation}
Here, $T_c$ is the contour ordering operator,  $\tau_{1,2}$ are the contour variables,
operators in the expression are in the Heisenberg picture, and $\langle\ldots\rangle=\mbox{Tr}[\ldots\hat\rho(t_0)]$.

Using (\ref{Snonint}) and Dyson equation for (\ref{Gdef}) leads to differential analog of the second law (\ref{law2})
in the form of {\em energy-resolved partial Clausius relation} (see Appendix~\ref{appA} for details)
\begin{equation}
\label{dlaw2}
\frac{d}{dt}S_B(t;E) =  \beta_B\,\dot{Q}_B(t;E) + \dot{P}_B(t;E)
\end{equation}
where 
\begin{align}
\label{SQPE_nonint}
& \frac{d}{dt}S_B(t;E) = \mbox{Tr}_S\bigg[\mathbf{i}^B(t;E)\,
\ln\frac{i\mathbf{G}^{>}(t,t)}{-i\mathbf{G}^{<}(t,t)}\bigg]
 \\
& \dot{Q}_B(t;E) =  (E-\mu_B)\,\mbox{Tr}_S\big[\mathbf{i}^B(t;E)\big]
 \nonumber \\
& \dot{P}_B(t;E) = \mbox{Tr}_S\bigg[\mathbf{i}^B(t;E)
\ln\frac{i\mathbf{G}^{>}(t,t)\, f_B(E)}{-i\mathbf{G}^{<}(t,t)\,[1-f_B(E)]}\bigg]
\nonumber
\end{align}
are the bath and energy resolved entropy, heat and entropy production fluxes, and where
\begin{align}
\label{iE}
 & i^B_{ij}(t;E) \equiv \sum_{n\in S}\int_{t_0}^t ds\,
 \\ &
 \bigg( 
 \big[\sigma^{B\, <}_{in}(t,s;E)\, G^{>}_{nj}(s,t) - \sigma^{B\, >}_{in}(t,s;E)\, G^{<}_{nj}(s,t)\big] e^{iE(s-t)}
\nonumber \\ &
 + \big[ G^{>}_{in}(t,s)\, \sigma^{B\, <}_{nj}(s,t;E) - G^{<}_{in}(t,s)\, \sigma^{B\, >}_{nj}(s,t;E) \big] e^{iE(t-s)}
 \bigg)
 \nonumber
\end{align}
is the matrix of energy-resolved particle flux at interface $S-B$. 
Here, $E$ is resolution in the energy of bath states, 
\begin{equation}
\begin{split}
\sigma^{B\, <}_{i_1i_2}(t_1,t_2;E) &\equiv i\, \gamma^B_{i_1i_2}(t_1,t_2;E)\, f_B(E)
\\
\sigma^{B\, >}_{i_1i_2}(t_1,t_2;E) &\equiv -i\, \gamma^B_{i_1i_2}(t_1,t_2;E)\, \big[1-f_B(E) \big]
\end{split}
\end{equation}
are the lesser and greater projections of the self-energy due to coupling to bath $B$,
$f_B(E)$ is the Fermi-Dirac thermal distribution, 
and~\cite{jauho_time-dependent_1994}
\begin{equation}
\gamma^B_{i_1i_2}(t_1,t_2;E) = 2\pi\sum_{k\in B} V_{i_1k}(t_1)\, V_{ki_2}(t_2)\, \delta(E-\varepsilon_k)
\end{equation}
is the energy-resolved dissipation matrix at interface $S-B$. 

Expressions (\ref{SQPE_nonint}) introduce energy resolved fluxes of entropy, heat and entropy production.
Corresponding total fluxes are obtained by summing over baths and integrating over energy
\begin{equation}
\label{SQP}
\begin{split}
\frac{d}{dt} S(t) &= \sum_B\int\frac{dE}{2\pi}\, \frac{d}{dt} S_B(t;E)
\\
\dot{Q}_B(t) &= \int\frac{dE}{2\pi}\, \dot{Q}_B(t;E)
\\
\dot{P}(t) &= \sum_B\int\frac{dE}{2\pi}\, \dot{P}_B(t;E)
\end{split}
\end{equation}
That is, in noninteracting systems the fluxes are exactly additive in terms of bath contributions
(below we show that the same is true in presence of interactions).
Thus, non-Markov character of Green's function formulation alleviates non-additivity problems 
of quantum master equation considerations~\cite{chan_quantum_2014,giusteri_interplay_2017,mitchison_non-additive_2018,friedman_quantum_2018}.

Additive form of fluxes, Eq.(\ref{SQP}), also indicates that for system which consists of several coupled parts,
\begin{equation}
\label{HNp}
\hat H_S(t) = \sum_{p=1}^{N_p} \hat H_S^{(p)}(t) + \sum_{p_1<p_2}^{N_p} \hat H_S^{(p_1p_2)}(t),
\end{equation} 
such that each part $p$ is connected to its own group of baths $\{B_p\}$,
one can use expressions for entropy of part $p$,
\begin{equation}
 S_p(t) = -\mbox{Tr}_p\big[\hat\rho_p(t)\,\ln\hat\rho_p(t)\big],
\end{equation}
and for multipartite mutual information,
\begin{equation}
\label{multi}
I_{1,\ldots,N_p}(t) = \sum_{p=1}^{N_p} S_p(t) - S(t),
\end{equation} 
to derive {\em energy resolved version of the local Clausius expression}
(see Appendix~\ref{appA} for details)
\begin{equation}
\label{Clausius_loc}
 \frac{d}{dt} S_p(t;E) = \sum_{B_p}\bigg( \beta_{B_p}\,\dot{Q}_{B_p}(t;E) + \dot{P}_{B_p}(t;E)\bigg) + \dot{I}_p(t;E)
\end{equation}
Here,
\begin{equation}
\label{dSpE_nonint}
\begin{split}
&\frac{d}{dt} S_p(t;E) = \sum_{B_p}\mbox{Tr}_p\bigg[\mathbf{i}_p^{B_p}(t;E)\, 
\ln\frac{i\mathbf{G}_{p}^{>}(t,t)}{-i\mathbf{G}_p^{<}(t,t)}\bigg] +
\\ &
2\pi\delta(E)\, 
\mbox{Tr}_p\bigg[ \big[ \mathbf{G}^{<}(t,t),\mathbf{H}^{(S)}(t)\big]_p\, 
 \ln\frac{i\mathbf{G}_{p}^{>}(t,t)}{-i\mathbf{G}_{p}^{<}(t,t)}\bigg]
 \end{split}
\end{equation}
and 
\begin{equation}
\label{dIpE_nonint}
\begin{split}
 \dot{I}_p(t;E) &\equiv \frac{d}{dt}S_p(t;E) - \sum_{B_p} \frac{d}{dt} S_{B_p}(t;E)
 \\ &
 = \dot{I}_p^{reg}(t;E) + \dot{I}_p^\delta(t;E)
 \\
 \dot{I}_p^{reg}(t;E) &\equiv \sum_{B_p}\bigg(
 \mbox{Tr}_p\bigg[\mathbf{i}_p^{B_p}(t;E)\, 
\ln\frac{i\mathbf{G}_{p}^{>}(t,t)}{-i\mathbf{G}_p^{<}(t,t)}\bigg]
\\ &\qquad\,\, 
- \mbox{Tr}_S\bigg[\mathbf{i}^{B_p}(t;E)\, 
\ln\frac{i\mathbf{G}^{>}(t,t)}{-i\mathbf{G}^{<}(t,t)}\bigg]
\bigg)
 \\
 \dot{I}_p^\delta(t;E) &\equiv 2\pi\delta(E)\times
 \\ &
  \mbox{Tr}_p\bigg[ \big[ \mathbf{G}^{<}(t,t),\mathbf{H}^{(S)}(t)\big]_p\, 
 \ln\frac{i\mathbf{G}_{p}^{>}(t,t)}{-i\mathbf{G}_{p}^{<}(t,t)}\bigg]
 \end{split}
\end{equation}
are the part $p$ entropy and information fluxes, respectively.
Subscript $p$ in $\big[ \mathbf{G}^{<}(t,t),\mathbf{H}^{(S)}(t)\big]_p$, 
$\mathbf{G}_{p}^{\gtrless}(t,t)$, and $\mathbf{i}_p^{B_p}(t;E)$ in (\ref{dSpE_nonint}) and (\ref{dIpE_nonint}) 
indicates matrices in the subspace $p$ of $S$.

\subsection{Interacting system}
We now turn to interacting systems. In the basis $\{\lvert S\rangle\}$ of many-body states of an isolated system 
contributions to the Hamiltonian (\ref{Hdef}) are
\begin{align}
\label{Hint}
 \hat H_S(t) &= \sum_{S_1,S_2\in S}  H^{S}_{S_1S_2}(t)\hat X_{S_1S_2}
\nonumber \\
 \hat H_B &= \sum_{k\in B} \varepsilon_k\hat c_k^\dagger\hat c_k
 \\
 \hat V_{SB}(t) &= \sum_{S_1,S_2\in S}\sum_{k\in B}
 \bigg( V_{(S_1S_2)k}(t)\,\hat X_{S_1S_2}^\dagger\hat c_k + H.c.\bigg)
 \nonumber
\end{align}
Here, $\hat X_{S_1S_2}\equiv \lvert S_1\rangle\langle S_2\rvert$ is the Hubbard operator.
Note that while focus of our consideration is on Fermi baths, generalization
to Bose case is straightforward.

Entropy of the system, Eq.~(\ref{Sdef}), is defined by the system density operator.
Its representation in the basis of the system many-body states is
\begin{equation}
 \big[\rho_S(t)\big]_{S_1S_2} = \langle\hat X_{S_2S_1}(t)\rangle
\end{equation}
Exact equation of motion (EOM) for the system density matrix is (see Appendix~\ref{appB} for details)
\begin{equation}
\label{XEOM}
\begin{split}
&\frac{d}{dt}\langle\hat X_{S_2S_1}(t)\rangle = -i \big[\mathbf{H}^{(S)}(t),\mathbf{\rho}_S(t)\big]_{S_1S_2}
\\ & + \sum_B \int\frac{dE}{2\pi}\, I^B_{S_1S_2}(t;E)
\end{split}
\end{equation}
where
\begin{align}
\label{ISSE}
 & I^B_{S_1S_2}(t;E) \equiv \sum_{S_3,S_4,S_5\in S} \int_{-\infty}^{t} ds
 \\ &
 \bigg(\ \Sigma^{B\, <}_{(S_3S_1)(S_4S_5)}(t,s;E)\, e^{iE(s-t)}\, \mathcal{G}^{>}_{(S_4S_5)(S_3S_2)}(s,t)
\nonumber \\ &
 - \Sigma^{B\, >}_{(S_3S_1)(S_4S_5)}(t,s;E)\, e^{iE(s-t)}\, \mathcal{G}^{<}_{(S_4S_5)(S_3S_2)}(s,t)
\nonumber \\ &
 - \Sigma^{B\, <}_{(S_2S_3)(S_4S_5)}(t,s;E)\, e^{iE(s-t)}\, \mathcal{G}^{>}_{(S_4S_5)(S_1S_3)}(s,t)
\nonumber \\ &
 + \Sigma^{B\, >}_{(S_2S_3)(S_4S_5)}(t,s;E)\, e^{iE(s-t)}\, \mathcal{G}^{<}_{(S_4S_5)(S_1S_3)}(s,t)
\nonumber \\ &
 + \mathcal{G}^{>}_{(S_3S_1)(S_4S_5)}(t,s)\, e^{iE(t-s)}\, \Sigma^{B\, <}_{(S_4S_5)(S_3S_2)}(s,t;E)
\nonumber \\ &
 - \mathcal{G}^{<}_{(S_3S_1)(S_4S_5)}(t,s)\, e^{iE(t-s)}\, \Sigma^{B\, >}_{(S_4S_5)(S_3S_2)}(s,t;E)
\nonumber \\ &
 - \mathcal{G}^{>}_{(S_2S_3)(S_4S_5)}(t,s)\, e^{iE(t-s)}\, \Sigma^{B\, <}_{(S_4S_5)(S_1S_3)}(s,t;E)
\nonumber \\ &
 + \mathcal{G}^{<}_{(S_2S_3)(S_4S_5)}(t,s)\, e^{iE(t-s)}\, \Sigma^{B\, >}_{(S_4S_5)(S_1S_3)}(s,t;E)
\bigg)
\nonumber
\end{align}
is the matrix of energy resolved probability flux at interface $S-B$, 
\begin{equation}
\begin{split}
\Sigma^{B\, <}_{(S_1S_3)(S_2S_4)}(t_1,t_2;E) &\equiv i\,\Gamma^B_{(S_1S_3)(S_2S_4)}(t_1,t_2; E)\, f_B(E)
\\
\Sigma^{B\, >}_{(S_1S_3)(S_2S_4)}(t_1,t_2;E) &\equiv -i\,\Gamma^B_{(S_1S_3)(S_2S_4)}(t_1,t_2; E)\, 
\\ &\qquad\qquad \times [1-f_B(E)]
\end{split}
\end{equation}
are the lesser and greater projections fo the self-energy due to coupling to bath $B$, $f_B(E)$ is the Fermi-Dirac
thermal distribution, and
\begin{equation}
\begin{split}
&\Gamma^B_{(S_1S_3)(S_2S_4)}(t_1,t_2;E)\equiv
\\ &\qquad
 2\pi\sum_{k\in B} V_{(S_1S_3)k}(t_1)\, V_{k(S_2S_4)}(t_2)\,
\delta(E-\varepsilon_k)
\end{split}
\end{equation}
is the energy-resolved dissipation matrix.
$\mathcal{G}^{\gtrless}_{(S_1S_3)(S_2S_4)}(t_1,t_2)$ are greater/lesser projections of the Hubbard Green's function
defined on the Keldysh contour as~\cite{chen_nonequilibrium_2017,miwa_towards_2017,bergmann_electron_2019}
\begin{equation}
\label{HubGdef}
\mathcal{G}_{(S_1S_3)(S_2S_4)}(\tau_1,\tau_2) \equiv -i\big\langle T_c\, \hat X_{S_1S_3}(\tau_1)
\hat X_{S_2S_4}^\dagger(\tau_2)\, \big\rangle
\end{equation}

EOM (\ref{XEOM}) leads to expression for second law in the form of energy-resolved version of the partial Clausius relation (\ref{dlaw2}) with entropy, heat and entropy production fluxes given by
(see Appendix~\ref{appC} for derivation)
\begin{align}
\label{SQPE}
& \frac{d}{dt} S_B(t;E) = -\mbox{Tr}_S\bigg[ \mathbf{I}^B(t;E)\, \ln \mathbf{\rho}_S(t) \bigg]
\\
& \dot{Q}_B(t;E) = (E-\mu_B) \mbox{Tr}_S\big[ \mathbf{N}\,\mathbf{I}^B(t;E) \big]
\nonumber \\
& \dot{P}_B(t;E) = -\mbox{Tr}_S\bigg[ \mathbf{I}^B(t;E)\bigg(\ln\mathbf{\rho}_S(t) - \mathbf{N}\ln\frac{f_B(E)}{1-f_B(E)}\bigg)\bigg]
\nonumber
\end{align}
Here, $\mathbf{N}$ is matrix representing the system number operator 
in the basis of many-body states $\{\lvert S\rangle\}$, its  elements 
$N_{S_1S_2}\equiv \langle S_1\rvert \hat N\lvert S_2\rangle = \delta_{N_{S_1},N_{S_2}}\, N_{S_1}$ 
yield information on number of electrons $N_S$ in state $\lvert S\rangle$.
As previously, total fluxes are given by integrating over energy and summing over baths, Eq.(\ref{SQP}). 

Finally, for the multipartite system (\ref{HNp}) one can derive
energy-resolved local Clausius expression (\ref{Clausius_loc}).
Derivation follows the same steps as in the noninteracting case considered in Appendix~\ref{appA}.
The only new feature is connection between probability and particle/energy fluxes as 
discussed in Appendix~\ref{appC}. 
Similar to noninteracting case, Eq.(\ref{dSpE_nonint}), 
consideration of part of the system results in non-zero contribution from
the first term in the right of (\ref{XEOM}) to the entropy flow
\begin{equation}
\label{dSpE}
\begin{split}
&\frac{d}{dt} S_p(t;E) = -\sum_{B_p}\mbox{Tr}_p\bigg[\mathbf{I}_p^{B_p}(t;E)\,\ln\mathbf{\rho}_p(t)\bigg]
\\&\qquad
+2\pi i\delta(E)\,
\mbox{Tr}_p\bigg[\big[\mathbf{H}^{(S)}(t),\mathbf{\rho}_S(t)\big]_p\,\ln\mathbf{\rho}_p(t)\bigg]
\end{split}
\end{equation}
Here, 
\begin{equation}
\begin{split}
&\hat\rho_p(t) \equiv \mbox{Tr}_{S\setminus p}\big[\hat\rho_S(t)\big],
\quad
\mathbf{I}_p^{B_p}(t;E) \equiv\mbox{Tr}_{S\setminus p}\big[\mathbf{I}^{B_p}(t;E)\big],
\\
&\big[\mathbf{H}^{(S)}(t),\mathbf{\rho}_S(t)\big]_p \equiv\mbox{Tr}_{S\setminus p}\bigg[ 
\big[\mathbf{H}^{(S)}(t),\mathbf{\rho}_S(t)\big]\bigg].
\end{split}
\end{equation}
Using (\ref{dlaw2}), (\ref{SQPE}) and (\ref{dSpE}) in energy resolved analog of (\ref{multi})
(see Eq.~(\ref{appA_multi}) in Appendix~\ref{appA}),
leads to the energy resolved local Clausius expression (\ref{Clausius_loc}) with
information flux given by
\begin{equation}
\label{dIpE}
\begin{split}
 \dot{I}_p(t;E) &= \dot{I}_p^{reg}(t;E) + \dot{I}_p^\delta(t;E)
 \\
 \dot{I}_p^{reg}(t;E) &\equiv \sum_{B_p}\bigg(
 \mbox{Tr}_S\bigg[\mathbf{I}^{B_p}(t;E)\, 
\ln\mathbf{\rho}_S(t)\bigg]
\\ &\qquad\,\, 
-  \mbox{Tr}_p\bigg[\mathbf{I}_p^{B_p}(t;E)\, 
\ln\mathbf{\rho}_{p}(t)\bigg]
\bigg)
 \\
 \dot{I}_p^\delta(t;E) &\equiv 2\pi i\delta(E)\times
 \\ &
\mbox{Tr}_p\bigg[\big[\mathbf{H}^{(S)}(t),\mathbf{\rho}_S(t)\big]_p\,\ln\mathbf{\rho}_p(t)\bigg] 
\end{split}
\end{equation}


\section{\label{numres}Numerical illustrations}
Here we illustrate our derivations with simulations performed in generic junction model sketched in
Fig.~\ref{fig1}. Fermi energy $E_F$ in each bath is taken as zero, and bias $V_{i}$ between left and right sides is applied
symmetrically, $\mu_{L_i}=E_F+V_i/2$ and $\mu_{R_i}=E_F-V_i/2$ ($i=1,2$).
Simulations are performed at steady state.

\begin{figure}[htbp]
\centering\includegraphics[width=\linewidth]{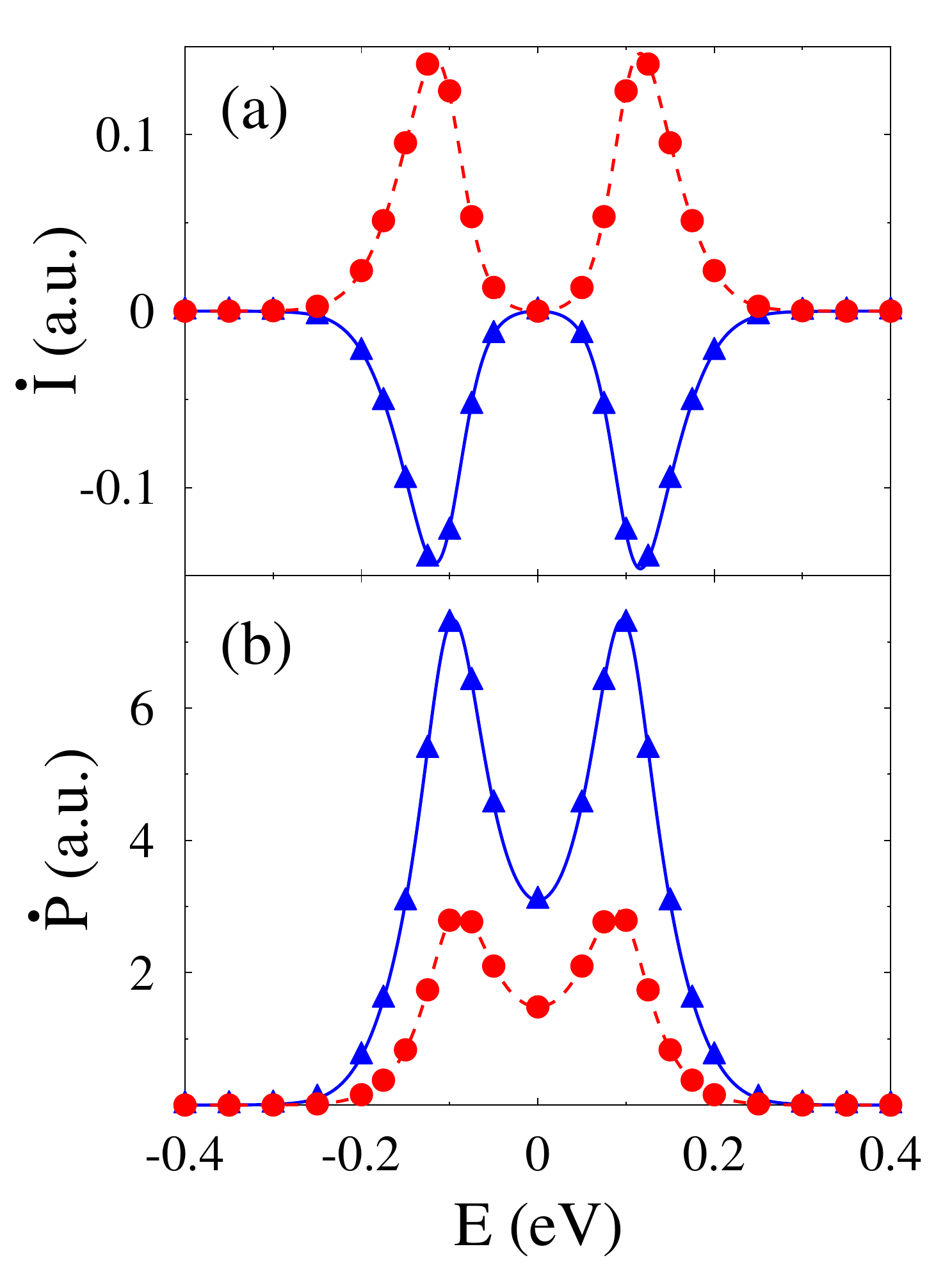}
\caption{\label{fig2}
Junction of Fig.~\ref{fig1} with $S$ represented by two-level system. 
Shown are steady state energy resolved (a) information flow $\dot{I}$, Eq.~({\ref{dIpE_nonint}}),
and (b) entropy production $\dot{P}$, Eq.~(\ref{SQPE_nonint}).
Characteristics of parts $S_1$ and $S_2$ are shown with solid (blue) and dashed (red) lines,
respectively. 
Triangles (blue) and circles (red) present information flow and entropy production evaluated
using formulation for interacting systems, Eqs.~(\ref{dIpE}) and (\ref{SQPE}). 
See text for parameters.
}
\end{figure}

First, we consider two-level system as an example for noninteracting $S$. 
Hamiltonian of the system is given in Eq.(\ref{HNp}) with  
\begin{equation}
\begin{split}
& \hat H_S^{(i)} = \varepsilon_i\hat d_i^\dagger\hat d_i\quad (i=1,2)
 \\
& \hat H_S^{(12)} = t_{12}\big(\hat d_1^\dagger\hat d_2+\hat d_2^\dagger\hat d_1\big)
 \end{split}
\end{equation}
Here, $\hat d_i^\dagger$ ($\hat d_i$) creates (annihilates) electron in level $\varepsilon_i$ of
subsystem $S_i$.

Figure~\ref{fig2} shows results of the simulation. Parameters are $T=300$~K, 
$\varepsilon_1=\varepsilon_2=0$, $t_{12}=0.1$~eV, $\gamma^B_{ij}=\delta_{i,j}\, 0.05$~eV,
$V_1=0.4$~eV and $V_2=-0.2$~eV. Simulation was performed on energy grid
spanning range from $-10$ to $10$~eV with step $0.001$~eV.
Panel (a) shows information flow from $S_1$ and into $S_2$, which is natural consequence of $V_1>\lvert V_2\rvert$.
Because of steady-state $\dot{I}_1=-\dot{I}_2$. 
Entropy production is presented in panel (b). The simulations were performed utilizing
non-interacting (lines) and interacting (markers) methodology. Fig.~\ref{fig2} shows close correspondence
between NEGF and Hubbard NEGF results.   

\begin{figure}[htbp]
\centering\includegraphics[width=\linewidth]{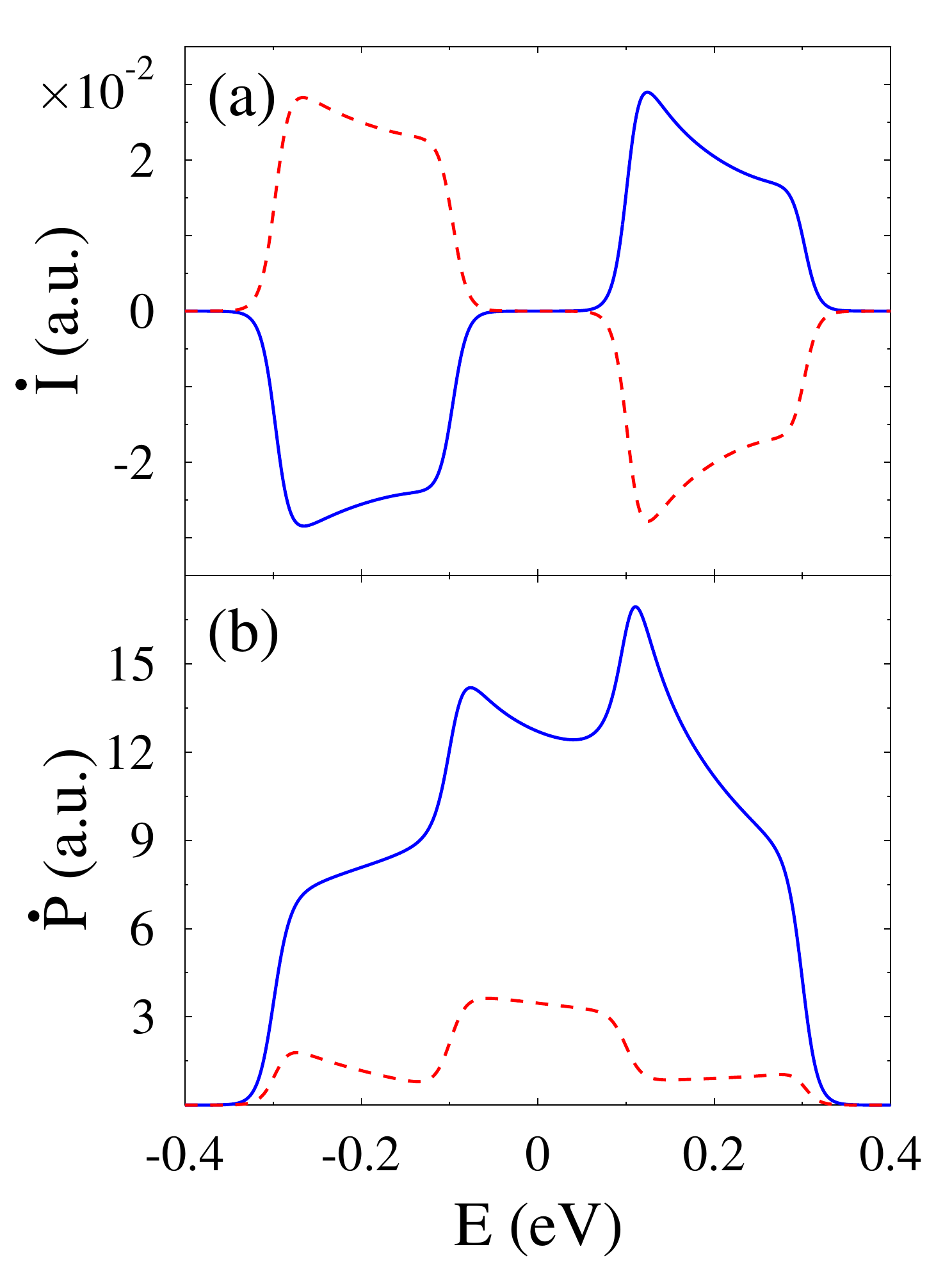}
\caption{\label{fig3}
Junction of Fig.~\ref{fig1} with $S$ represented by quantum dot in the limit of strong Coulomb repulsion.
Subsystem $S_1$ represents spin up  and $S_2$ represents spin down levels.
Shown are steady state energy resolved (a) information flow $\dot{I}$, Eq.~({\ref{dIpE}}),
and (b) entropy production $\dot{P}$, Eq.~(\ref{SQPE}).
Characteristics of parts $S_1$ and $S_2$ are shown with solid (blue) and dashed (red) lines,
respectively. See text for parameters.
}
\end{figure}

We now turn to consider interacting system.
As a toy model we take quantum dot junction in the regime of strong Coulomb repulsion.
Subsystems $S_1$ and $S_2$ represent spin up and spin down levels of the system. 
The dot is subjected coupling to external spin degree of freedom (e.g. magnetic adatom)
induces spin-flip processes which we model by effective electron transfer matrix element between the two levels. 
Hamiltonian has a form (\ref{HNp}) with
\begin{equation}
\begin{split}
& \sum_{i=1}^2 \hat H_S^{(i)}= \sum_{\sigma=\uparrow,\downarrow} \varepsilon_{\sigma}\hat d_{\sigma}^\dagger\hat d_{\sigma} + U\hat n_{\uparrow}\hat n_{\downarrow}
 \\
 & \hat H_S^{(12)} = t_{12}\left(\hat d_\uparrow^\dagger\hat d_{\downarrow} +H.c.\right)
 \end{split}
\end{equation}
Here, $\hat d_{\sigma}^\dagger$ ($\hat d_{\sigma}$) creates (annihilates) electron of spin $\sigma$
in level  $\varepsilon_{\sigma}$, $\hat n_{\sigma}\equiv\hat d_{\sigma}^\dagger\hat d_{\sigma}$.
We consider strong Coulomb repulsion, $U\to \infty$, in the parameter regime where Kondo feature
is detectable. Simulations were performed utilizing equation of motion approach to the Hubbard
Green's functions within a method presented in Ref.~\onlinecite{ochoa_non-equilibrium_2014}.
We note that such consideration yields Kondo feature only qualitatively, which is enough for
illustration purposes. More elaborate considerations should employ more advanced decoupling 
scheme of Ref.~\onlinecite{lacroix_density_1981} or rely on numerically exact techniques.

Figure~\ref{fig3} shows results of the simulation. Parameters are $T=100$~K, 
$\varepsilon_\uparrow=\varepsilon_\downarrow=-2$~eV, $t_{12}=0.5$~eV, $\gamma^B_{ij}=\delta_{i,j}\, 0.5$~eV,
$V_\uparrow=0.6$~eV and $V_\downarrow=0.2$~eV. Simulation was performed on energy grid
spanning range from $-10$ to $10$~eV with step $0.001$~eV.
Information flow (panel a), entropy production is presented in panel (b).
Note that entropy production for $S_1$ demonstrate Kondo features near chemical potentials of 
the opposite spin ($E=\pm 0.1$~eV). The feature at $E=0.1$~eV is also observed in the information flow. 
Note also that contrary to the noninteracting case, where information flow is always 
from $S_1$ into $S_2$, interacting model yields more structured energy profile.
This example indicates potential importance of strong correlations in thermodynamic properties of
systems strongly coupled to their baths.

\section{\label{conclude}Conclusion}
Starting from von Neumann expression with reduced (system) density matrix as definition of thermodynamic entropy
for a system strongly coupled to its baths, we formulate the second law of thermodynamics in the form of
bath and energy resolved partial and local Clausius relations. The derivation yields expressions
for entropy, heat, and information flows and for entropy production rate formulated in terms
of system variables only. This makes the expressions suitable for actual calculations of the thermodynamic 
characteristics of open nanoscale molecular devices. 

The expressions are derived using the standard NEGF for noninteracting and Hubbard NEGF
for interacting systems. Utilization of Green's function techniques allows to generalize Markov weak coupling
(second order in system-baths) consideration of Ref.~\cite{ptaszynski_thermodynamics_2019}
to strongly coupled case of non-Markov system dynamics. 
In particular, we utilize exact equation of motion for the reduced (system) density matrix
in  terms of Green's functions, introduce energy resolution of the thermodynamic flows,
and show that Green's function analysis allows to alleviate non-separability problems 
of density matrix considerations and thus to introduce exact general form of information flow.
Expressions for thermodynamic characteristics reduce to known expected limiting forms 
in weak coupling or steady-state situations.
We stress that while theoretical consideration in the manuscript focuses on Fermi baths
and charge flux between the system and baths, generalization to incorporate Bose degrees of freedom 
and energy flux is straightforward within the interacting (Hubbard NEGF) formulation.

Theoretical derivations are illustrated with numerical results for a junction
which consists fo two sub-systems each coupled its own pair of baths (Fig.~\ref{fig1}).
Generic models of the two-level system and quantum dot with spin-flip processes 
are used as illustrations for noninteracting and interacting system, respectively. 
Energy resolved thermodynamic characteristics of the interacting system are shown
to demonstrate strongly correlated features (Kondo), which in principle can be detected
in corresponding flows similar to conductance measurements.

\begin{acknowledgments}
We acknowledge useful discussions with Massimiliano Esposito and Raam Uzdin.
\end{acknowledgments}
\appendix
\section{Derivation of energy resolved partial and local Clausius expression for noninteracting system}\label{appA}
For noninteracting systems (i.e. systems described by quadratic Hamiltonian), the Wick's theorem 
holds for any part of the universe (system plus baths). This means that corresponding many-body 
density operator for any part of the system has Gaussian form.
Thus, path integral formulation yields for the local density matrix~\cite{bergmann_greens_2020}
\begin{equation}
\label{appA_rhoA}
 \rho_A(t) = \mbox{det}\big[i\mathbf{G}_{A}^{>}(t,t)\big]\frac{-i\mathbf{G}_A^{<}(t,t)}{i\mathbf{G}_A^{>}(t,t)}
\end{equation}
Here, $A$ can be any part of the universe. Below we consider system ($A=S$) or its part ($A=p$),
and assume that each part $p$ of the system is coupled to its own set of baths $\{ B_p\}$.

Assuming von Neumann expression for entropy of the part $A$, Eq.(\ref{appA_rhoA}) leads to 
\begin{equation}
\label{appA_SA}
\begin{split}
 S_A(t) =& -\mbox{Tr}_A\big[\rho_A(t)\,\ln\rho_A(t)\big] 
 \\  =& 
  -\mbox{Tr}_A\big[-i\mathbf{G}_A^{<}(t,t)\,\ln\big(-i\mathbf{G}_A^{<}(t,t)\big)\big]
  \\ &
  -\mbox{Tr}_A\big[i\mathbf{G}_A^{>}(t,t)\,\ln\big(i\mathbf{G}_A^{>}(t,t)\big)\big]
  \end{split}
\end{equation}
Here, $\mathbf{G}_A^{\gtrless}(t,t)$ is the greater/lesser projection of the single-particle 
Green's function for the system, $\mathbf{G}^{\gtrless}(t,t)$ ($A=S$), or its part,
$\mathbf{G}^{\gtrless}_p(t,t)$ ($A=p$).
Taking time derivative of (\ref{appA_SA}) leads to expression for entropy flux
\begin{equation}
\label{appA_dSA_1}
\frac{d}{dt} S_A(t) = \mbox{Tr}_A\bigg[\bigg(-i\frac{d}{dt} \mathbf{G}_A^{<}(t,t)\bigg)\, 
\ln\frac{i\mathbf{G}_{A}^{>}(t,t)}{-i\mathbf{G}_A^{<}(t,t)}\bigg]
\end{equation}

Dyson equation for the lesser projection of the single particle Green's function is
\begin{align}
\label{appA_EOM_1}
 &-i \frac{d}{dt} \big(G_A^{<}(t,t)\big)_{ij} =
 \big[ \mathbf{G}^{<}(t,t) , \mathbf{H}^{(S)}(t) \big]_{ij}
 \\ &
 +\sum_{n\in S}\int_{t_0}^t ds\, \bigg( G^{r}_{in}(t,s)\,\sigma^{<}_{nj}(s,t) + G^{<}_{in}(t,s)\,\sigma^{a}_{nj}(s,t)
 \nonumber\\ &\qquad\qquad\quad
 - \sigma^{r}_{in}(t,s)\, G^{<}_{nj}(s,t)  - \sigma^{<}_{in}(t,s)\, G^{a}_{nj}(s,t) \bigg)
\nonumber
\end{align}
Here, $i,j\in A$,  $</r/a$ is the lesser/retarded/advanced projection, $G$ is defined in (\ref{Gdef}),
and for noninteracting system self-energy $\sigma$ is 
\begin{equation}
\label{appA_sigma}
\begin{split}
\sigma_{ij}(\tau,\tau') &= \sum_B \sigma^B_{ij}(\tau,\tau')
\\
\sigma^B_{ij}(\tau,\tau') &\equiv \sum_{k\in B} V_{ik}(t)\, g_k(\tau,\tau')\, V_{kj}(t')
\end{split}
\end{equation} 
where $g_k(\tau,\tau')$ is single-particle Green's function of free electron in state $k$ of the bath $B$.
Substituting (\ref{appA_sigma}) into (\ref{appA_EOM_1}) and using
\begin{equation}
\begin{split}
\mathbf{G}^r(t,t') &= \theta(t-t')\big( \mathbf{G}^{>}(t,t') - \mathbf{G}^{<}(t,t') \big)
\\
\mathbf{G}^a(t,t') &= \theta(t'-t)\big( \mathbf{G}^{<}(t,t') - \mathbf{G}^{>}(t,t') \big)
\end{split}
\end{equation}
and similar expressions for $\sigma$ leads to
\begin{align}
\label{appA_EOM_2}
& -i \frac{d}{dt} \big(G_A^{<}(t,t)\big)_{ij} =
 \big[ \mathbf{G}^{<}(t,t) , \mathbf{H}^{(S)}(t) \big]_{ij}
 \\ &
 +\sum_B\sum_{n\in S}\int_{t_0}^t ds\, \big( 
 \sigma^{B\, <}_{in}(t,s)\, G^{>}_{nj}(s,t)  + G^{>}_{in}(t,s)\,\sigma^{B\, <}_{nj}(s,t)
 \nonumber\\ &\qquad\qquad\qquad
 - \sigma^{B\, >}_{in}(t,s)\, G^{<}_{nj}(s,t) - G^{<}_{in}(t,s)\,\sigma^{B\, >}_{nj}(s,t)
  \big)
  \nonumber
\end{align}
Using the Jauho-Wingreen-Meir expression for particle flux $I_B(t)$ at interface $S-B$~\cite{jauho_time-dependent_1994}
\begin{equation}
\begin{split}
&I_B(t) = 
\\ &
\int_{t_0}^t ds\, \mbox{Tr}_S\bigg[
 \mathbf{\sigma}^{B\, <}(t,s)\, \mathbf{G}^{>}(s,t)  + \mathbf{G}^{>}(t,s)\,\mathbf{\sigma}^{B\, <}(s,t)
 \\ &\qquad\qquad
-\mathbf{\sigma}^{B\, >}(t,s)\, \mathbf{G}^{<}(s,t) - \mathbf{G}^{<}(t,s)\,\mathbf{\sigma}^{B\, >}(s,t)
\bigg],
\end{split}
\end{equation}
(\ref{appA_sigma}) with
\begin{equation}
\label{appA_gkltgt}
\begin{split}
 g_k^{<}(t,t') &= i n_k e^{-i\varepsilon_k(t-t')}
 \\
 g_k^{>}(t,t') &= -i [1-n_k] e^{-i\varepsilon_k(t-t')},
\end{split}
\end{equation}
and
\begin{equation}
\label{appA_sum2int}
 \sum_{k\in B} \ldots\equiv \int \frac{dE}{2\pi}\, 2\pi\sum_{k\in B}\delta(E-\varepsilon_k)\ldots
\end{equation}
leads to
\begin{equation}
\label{appA_EOM_3}
\begin{split}
 -i \frac{d}{dt} \big(G_A^{<}(t,t)\big)_{ij} &=
 \big[ \mathbf{G}^{<}(t,t) , \mathbf{H}^{(S)}(t) \big]_{ij}
 \\ &
 +\sum_{B_A}\int\frac{dE}{2\pi}\, i^{B_A}_{ij}(t;E)
 \end{split}
\end{equation}
where the matrix of energy-resolved particle flux $\mathbf{i}^B(t;E)$ is defined in (\ref{iE}).

Substituting (\ref{appA_EOM_3}) into (\ref{appA_dSA_1}) yields
\begin{equation}
\label{appA_dSA_2}
\begin{split}
\frac{d}{dt} S_A(t) &= 
 \mbox{Tr}_A\bigg[ \big[ \mathbf{G}^{<}(t,t),\mathbf{H}^{(S)}(t)\big]_A\, 
 \ln\frac{i\mathbf{G}_{A}^{>}(t,t)}{-i\mathbf{G}_A^{<}(t,t)}\bigg]
 \\ &
 + \sum_{B_A}\int\frac{dE}{2\pi}\,\mbox{Tr}_A\bigg[\mathbf{i}_A^{B_A}\, 
\ln\frac{i\mathbf{G}_{A}^{>}(t,t)}{-i\mathbf{G}_A^{<}(t,t)}\bigg]
\end{split}
\end{equation}
where subscript $A$ in $\big[ \mathbf{G}^{<}(t,t),\mathbf{H}^{(S)}(t)\big]_A$, $\mathbf{G}_A^{\gtrless}(t,t)$,
and $\mathbf{i}_A^{B_A}$ indicates sub-matrix with indices in subspace $A$.

\subsection*{Partial Clausius expression, Eq.(\ref{dlaw2})}
We now restrict our consideration to the system ($A=S$).
In this case first term in the right side of (\ref{appA_dSA_2}) is identically zero,
and we can introduce energy-resolved version of the entropy flux as defined in (\ref{SQPE_nonint}) 
and (\ref{SQP}). Energy resolved version of the partial Clausius inequality (\ref{dlaw2})
follows from 
\begin{equation}
 \ln\frac{f_B(E)}{1-f_B(E)} = -\beta_B(E-\mu_B)
\end{equation}
and definitions of energy resolved heat and entropy production fluxes given in (\ref{SQPE_nonint}).
Note that definition of the heat flux $\dot{Q}_B(t)$ is consistent with the standard quantum transport definitions of
particle $I_B(t)$ and energy $J_B(t)$ fluxes
\begin{equation}
\label{appA_QJIdef}
\begin{split}
\dot{Q}_B(t) &\equiv J_B(t) -\mu_B\, I_B(t)
\\
J_B(t) &\equiv -\frac{d}{dt} \langle\hat H_B(t)\rangle
\\
I_B(t) &\equiv -\frac{d}{dt}\sum_{k\in B}\langle\hat c_k^\dagger(t)\hat c_k(t)\rangle
\end{split}
\end{equation}

\subsection*{Local Clausius expression, Eq.(\ref{Clausius_loc})}
Now we restrict our consideration to part $p$ of the system ($A=p$).
Part $p$ is assumed to be coupled to its own set of baths $\{B_p\}$, so that $\sum_B\sum_{n\in S}\ldots$ in
the second term in the right of (\ref{appA_EOM_2}) is substituted by $\sum_{B_p}\sum_{n\in p}\ldots$.
Thus, matrix for energy-resolved particle $i^B_{ij}$, Eq.~(\ref{iE}), is substituted by 
similar expression $\big[i_p^{B_p}\big]_{ij}$ with $i,j\in p$ and $\sum_{n\in S}\ldots\to\sum_{n\in p}\ldots$
\begin{align}
\label{appB_ipE}
 & \big[i_p^{B_p}\big]_{ij}(t;E) \equiv \sum_{n\in p}\int_{t_0}^t ds\,
 \\ &
 \bigg( 
 \big[\sigma^{B_p\, <}_{in}(t,s;E)\, G^{>}_{nj}(s,t) - \sigma^{B_p\, >}_{in}(t,s;E)\, G^{<}_{nj}(s,t)\big] e^{iE(s-t)}
\nonumber \\ &
 + \big[ G^{>}_{in}(t,s)\, \sigma^{B_p\, <}_{nj}(s,t;E) - G^{<}_{in}(t,s)\, \sigma^{B_p\, >}_{nj}(s,t;E) \big] e^{iE(t-s)}
 \bigg)
 \nonumber
\end{align}
This is the current expression in the second term of (\ref{appA_dSA_2}).

Main difference (when compared with $A=S$) is non-zero contribution from the first term in the right
in (\ref{appA_dSA_2}). 
Thus, energy-resolved expression for bath specific entropy flux of the system, Eq.~(\ref{SQPE_nonint}),
is substituted with part specific expression, Eq.(\ref{dSpE_nonint}), with additional delta-term in it.

Using (\ref{dlaw2}), (\ref{SQPE_nonint}) and (\ref{dSpE_nonint}) in energy resolved analog of (\ref{multi}),
\begin{equation}
\label{appA_multi}
 I_{1,\ldots,N_p} =\int\frac{dE}{2\pi}\, I_{1,\ldots,N_p}(t;E),
\end{equation}
leads to separability of the multipartite mutual information into part-specific contributions
\begin{equation}
\begin{split}
&\dot{I}_{1,\ldots,N_p}(t;E) = \sum_p \frac{d}{dt} S_p(t;E) 
\\ &\quad
- 
\sum_p\sum_{B_p}\bigg(\beta_{B_p}\dot{Q}_{B_p}(t;E) + \dot{P}_{B_p}(t;E)\bigg) 
\\ &\quad
\equiv \sum_p \dot{I}_p(t;E)
\end{split}
\end{equation}
Rearranging terms in the latter expression yields (\ref{Clausius_loc}) with
information flux consisting of regular and delta-type contributions as defined in Eq.(\ref{dIpE_nonint}).

\section{Derivation of exact EOM for system density matrix, Eq.(\ref{XEOM})}\label{appB}
We start with writing Heisenberg EOM for $\hat X_{S_2S_1}(t)$ under dynamics governed by the Hamiltonian
(\ref{Hdef}) and (\ref{Hint}). Averaging result over an initial density operator leads to
\begin{align}
\label{appB_XEOM_1}
& \frac{d}{dt} \big\langle \hat X_{S_2S_1}(t)\big\rangle =
\nonumber \\ &
 i\sum_{S_3}\big\langle H^{(S)}_{S_3S_2}(t)\,\hat X_{S_3S_1}(t) - H^{(S)}_{S_1S_3}(t)\, \hat X_{S_2S_3}(t) \big\rangle
 \nonumber \\ &
 + \sum_B\sum_{k\in B}\sum_{S_3\in S}
 \\ &
 \bigg(\ \ 
 V_{(S_2S_3)\, k}(t)\,\mathcal{G}^{<}_{k\, (S_1S_3)}(t,t) - V_{(S_3S_1)\, k}(t)\,\mathcal{G}^{<}_{k\,(S_3S_2)}(t,t)
\nonumber \\ &
 + V_{k\,(S_3S_2)}(t)\,\mathcal{G}^{<}_{(S_3S_1)\, k}(t,t) - V_{k\,(S_1S_3)}(t)\,\mathcal{G}^{<}_{(S_2S_3)\, k}(t,t)
 \bigg)
 \nonumber
\end{align}
where $\mathcal{G}_{(SS')\, k}^{\gtrless}(t,t)$ and $\mathcal{G}^{\gtrless}_{k\,(SS')}(t,t)$
are the equal time greater/lesser projections of the Hubbard Green's functions defined on the contour as
\begin{equation}
\label{appB_GSSkdef}
\begin{split}
\mathcal{G}_{(SS')\, k}(\tau,\tau') &\equiv
 -i\big\langle T_c\,\hat X_{SS'}(\tau)\,\hat c_k^\dagger(\tau')\big\rangle
\\
\mathcal{G}_{k\,(SS')}(\tau,\tau') &\equiv 
-i\big\langle T_c\,\hat c_k(\tau)\,\hat X_{SS'}^\dagger(\tau')\big\rangle
\end{split}
\end{equation}
Integral forms of the right and left Dyson equations for (\ref{appB_GSSkdef}) are
\begin{align}
\label{appB_GSSkEOM}
 &\mathcal{G}_{(SS')\, k}(\tau,\tau') = 
 \\ &
 \sum_{S_4,S_5\in S}\int_c d\sigma\,
 \mathcal{G}_{(SS')(S_4S_5)}(\tau,\sigma)\, V_{(S_4S_5)\, k}(s)\, g_k(\sigma,\tau') 
\nonumber \\
&\mathcal{G}_{k\,(SS')}(\tau,\tau') = 
\nonumber \\ &
\sum_{S_4,S_5\in S}\int_c d\sigma\,
 g_{k}(\tau,\sigma)\, V_{k\,(S_4S_5)}(s)\,\mathcal{G}_{(S_4S_5)(SS')}(\sigma,\tau')  
\nonumber
\end{align}
where 
\begin{equation}
 g_k(\tau,\tau')\equiv -i \langle T_c\,\hat c_k(\tau)\,\hat c_k^\dagger(\tau')\rangle
\end{equation}
is the single-particle Green's function of free electron in state $k$ of contact $B$.

\begin{figure}[htbp]
\centering\includegraphics[width=\linewidth]{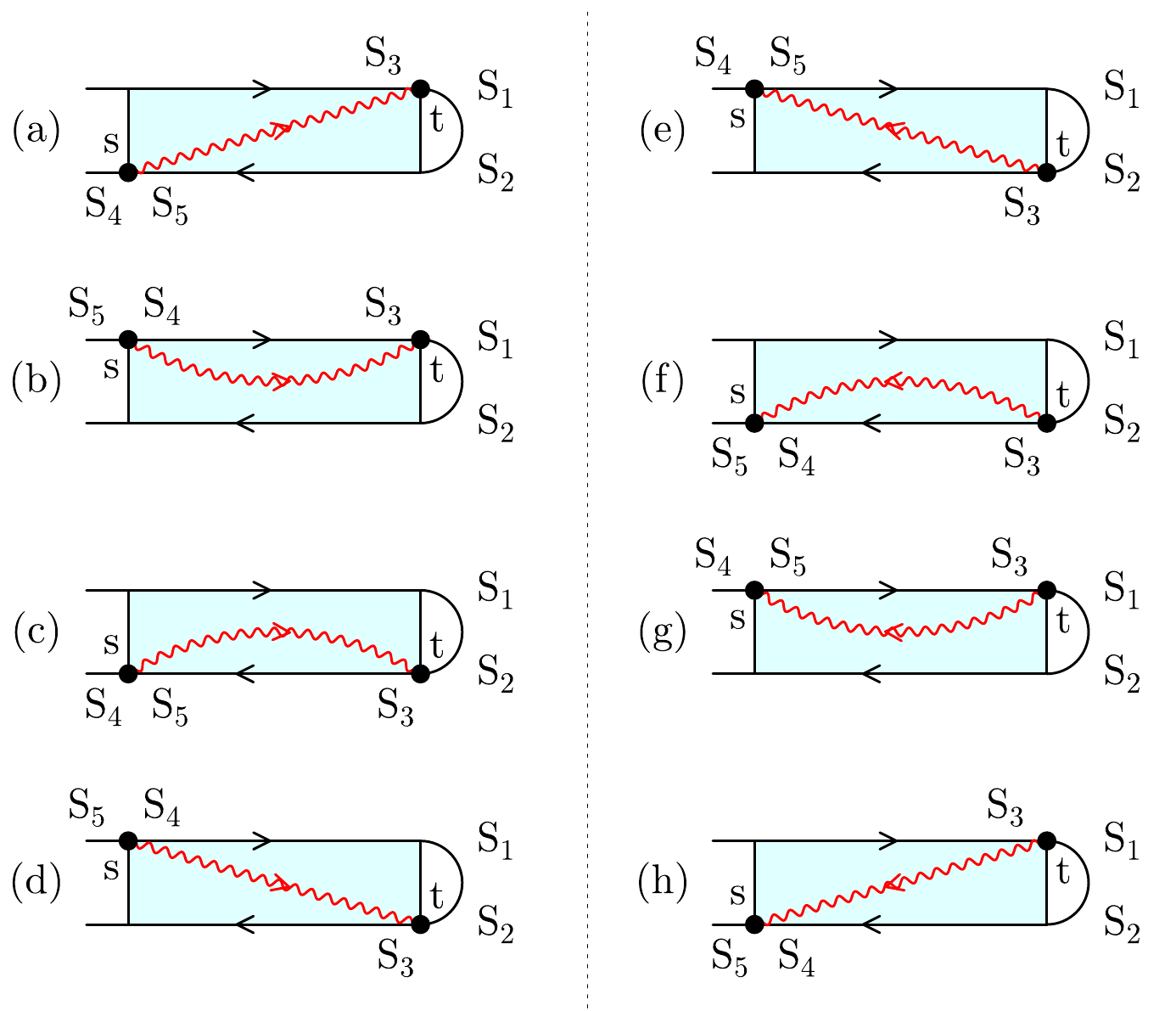}
\caption{\label{fig_appB}
Projections representing terms in Eq.~(\ref{appB_XEOM_2}).
Directed wavy line (red) indicates self-energy, shaded square (cyan) represents evolution 
accounted for by the Hubbard Green's function.
}
\end{figure}

Substituting lesser projections of (\ref{appB_GSSkEOM}) into (\ref{appB_XEOM_1}) leads to
\begin{align}
\label{appB_XEOM_2}
& \frac{d}{dt} \big\langle \hat X_{S_2S_1}(t)\big\rangle = -i \big[\mathbf{H}^{(S)}(t),\mathbf{\rho}_S(t)\big]_{S_1S_2}
\\ &
 +\sum_B\sum_{S_3,S_4,S_5\in S} \int_{t_0}^{t} ds
\nonumber \\ &
 \bigg(\ \Sigma^{B\, <}_{(S_3S_1)(S_4S_5)}(t,s)\, \mathcal{G}^{>}_{(S_4S_5)(S_3S_2)}(s,t) & \mbox{(a)}
\nonumber \\ &
 - \Sigma^{B\, >}_{(S_3S_1)(S_4S_5)}(t,s)\,  \mathcal{G}^{<}_{(S_4S_5)(S_3S_2)}(s,t) & \mbox{(b)}
\nonumber \\ &
 - \Sigma^{B\, <}_{(S_2S_3)(S_4S_5)}(t,s)\,  \mathcal{G}^{>}_{(S_4S_5)(S_1S_3)}(s,t) & \mbox{(c)}
\nonumber \\ &
 + \Sigma^{B\, >}_{(S_2S_3)(S_4S_5)}(t,s)\,  \mathcal{G}^{<}_{(S_4S_5)(S_1S_3)}(s,t) & \mbox{(d)}
\nonumber \\ &
 + \mathcal{G}^{>}_{(S_3S_1)(S_4S_5)}(t,s)\, \Sigma^{B\, <}_{(S_4S_5)(S_3S_2)}(s,t) & \mbox{(e)}
\nonumber \\ &
 - \mathcal{G}^{<}_{(S_3S_1)(S_4S_5)}(t,s)\, \Sigma^{B\, >}_{(S_4S_5)(S_3S_2)}(s,t) & \mbox{(f)}
\nonumber \\ &
 - \mathcal{G}^{>}_{(S_2S_3)(S_4S_5)}(t,s)\, \Sigma^{B\, <}_{(S_4S_5)(S_1S_3)}(s,t) & \mbox{(g)}
\nonumber \\ &
 + \mathcal{G}^{<}_{(S_2S_3)(S_4S_5)}(t,s)\, \Sigma^{B\, >}_{(S_4S_5)(S_1S_3)}(s,t) \bigg) & \mbox{(h)}
\nonumber
\end{align}
where $\Sigma^{B\,\gtrless}_{(S_1S_2)(S_3S_4)}(t,t')$ are the greater/lesser 
projections of the self-energy due to coupling to contact $B$
\begin{equation}
\begin{split}
& \Sigma^B_{(S_1S_2)(S_3S_4)}(\tau,\tau') \equiv 
\\ &\quad
\sum_{k\in B} V_{(S_1S_2)\, k}(t)\, g_k(\tau,\tau')\,
V_{k\, (S_3S_4)}(t')
\end{split}
\end{equation}
Graphic representation of the projections in (\ref{appB_XEOM_2}) is given in Fig.~\ref{fig_appB}.
We note in passing that suggestion of Ref.~\onlinecite{whitney_non-markovian_2018} to drop same branch projections 
(e.g., projections (b), (f), (c) and (g) in second order in system-bath coupling) when
building thermodynamic description is inconsistent with assumption of von Neumann form of the
system entropy. 
Note also that connection between exact EOM (\ref{appB_XEOM_2}) and approximate 
quantum master equations for reduced density matrix was discussed in Ref.~\onlinecite{bergmann_electron_2019}.

Finally, using (\ref{appA_gkltgt}) and (\ref{appA_sum2int}) in (\ref{appB_XEOM_2}) leads to (\ref{XEOM}).

\section{Derivation of partial Clausius expression for interacting system}\label{appC}
We start from von Neumann expression for system entropy, Eq.~(\ref{Sdef}),
and take time derivative to get entropy flux
\begin{equation}
\label{appC_dS}
\frac{d}{dt} S(t) = -\mbox{Tr}_S\bigg[\frac{d}{dt} \mathbf{\rho}_S(t)\,\ln\mathbf{\rho}_S(t)\bigg]
\end{equation}
Using exact EOM for the system density matrix, Eq.(\ref{XEOM}), and taking into account that 
first term in the right of (\ref{XEOM}) gives zero contribution to the trace in (\ref{appC_dS})
and second term in the right of (\ref{XEOM}) is exactly separable into energy-resolved contributions from different baths
leads to
\begin{equation}
 \frac{d}{dt} S(t) =\sum_B\int\frac{dE}{2\pi}\, \frac{d}{dt} S_B(t;E)
\end{equation}
where energy-resolved entropy flux $\frac{d}{dt} S_B(t;E)$ is defined in (\ref{SQPE}).

Contrary to noninteracting case, second term in the right of (\ref{XEOM})
is not directly related to particle flux, rather it is probability flux.
Indeed, writing Heisenberg EOM for the current at interface $S-B$, Eq.~(\ref{appA_QJIdef}), yields
\begin{equation}
\begin{split}
I_B(t) &\equiv -\frac{d}{dt}\sum_{k\in B} \big\langle\hat c_k^\dagger(t)\hat c_k(t)\big\rangle
\\ &
= -i\sum_{k\in B}\big\langle\big[\hat H(t),\hat c_k^\dagger(t)\hat c_k(t)\big]\big\rangle
\\ &
=-2\,\mbox{Re}\sum_{S_1,S_2\in S}\sum_{k\in B} V_{(S_1S_2)\, k}(t)\,\mathcal{G}^{<}_{k\,(S_1S_2)}(t,t)
\\ &
= 2\mbox{Re}\sum_{S_1,S_2,S_3,S_4\in S} \int_{t_0}^t ds\,
\\ &
\bigg(\,\,
\Sigma^{B\, <}_{(S_1S_2)(S_3S_4)}(t,s)\, \mathcal{G}^{>}_{(S_3S_4)(S_1S_2)}(s,t)
\\ &
- \Sigma^{B\, >}_{(S_1S_2)(S_3S_4)}(t,s)\, \mathcal{G}^{<}_{(S_3S_4)(S_1S_2)}(s,t)
\bigg)
\end{split}
\end{equation}
which is not equivalent to expression in the right of (\ref{XEOM}).

Connection between probability and particle fluxes is established considering 
change in particle number of the system caused by coupling to baths
\begin{equation}
\begin{split}
 \frac{d}{dt}\langle \hat N(t)\rangle &= \mbox{Tr}_S\bigg[\hat N\,\frac{d}{dt}\hat\rho_S(t)\bigg]
 \\ &
 = \sum_{S_1\in S}\sum_B\int\frac{dE}{2\pi}\, N_{S_1}\, I^B_{S_1S_1}(t;E)
 \end{split}
\end{equation}
where energy resolved probability flux is defined in (\ref{ISSE}).
Thus, energy-resolved particle flux at interface $S-B$ is
\begin{equation}
\label{appC_particle}
 \sum_{S_1\in S} N_{S_1}\, I^B_{S_1S_1}(t;E)
\end{equation}
Similarly, energy-resolved energy flux at interface $S-B$ is
\begin{equation}
\label{appC_energy}
 \sum_{S_1\in S} E\, N_{S_1}\, I^B_{S_1S_1}(t;E)
\end{equation}
Expressions (\ref{appC_particle}) and (\ref{appC_energy}) are used in definition of 
heat flux in (\ref{SQPE}). 

Energy resolved partial Clausius expression (\ref{dlaw2}) for interacting system 
directly follows from (\ref{appC_dS}) and definitions in (\ref{SQPE}).


\end{document}